\documentclass[DIV=calc, paper=a4, fontsize=11pt, twocolumn]{scrartcl}	 
\usepackage[english]{babel} 
\usepackage[protrusion=true,expansion=true]{microtype} 
\usepackage{amsmath,amsfonts,amsthm} 
\usepackage[svgnames]{xcolor} 
\usepackage[hang, small,labelfont=bf,up,textfont=it,up]{caption} 
\usepackage{booktabs} 
\usepackage{fix-cm}	 
\usepackage{graphicx}

\usepackage{sectsty} 
\allsectionsfont{\usefont{OT1}{phv}{b}{n}} 

\usepackage{fancyhdr} 
\usepackage{lastpage} 

\usepackage{lettrine} 
\newcommand{\initial}[1]{ 
\lettrine[lines=3,lhang=0.3,nindent=0em]{
\color{DarkGoldenrod}
{\textsf{#1}}}{}}
\usepackage{cite}

\usepackage{titling} 

\newcommand{\HorRule}{\color{DarkGoldenrod} \rule{\linewidth}{1pt}} 

\pretitle{\vspace{-30pt} \begin{flushleft} \HorRule \fontsize{50}{50} \usefont{OT1}{phv}{b}{n} \color{DarkRed} \selectfont} 

\title{Pornography consumption in Social Media} 

\posttitle{\par\end{flushleft}\vskip 0.5em} 

\preauthor{\begin{flushleft}\large \lineskip 0.5em \usefont{OT1}{phv}{b}{sl} \color{DarkRed}} 

\author{Mauro Coletto, Luca Maria Aiello, Claudio Lucchese, Fabrizio Silvestri } 

\postauthor{\footnotesize \usefont{OT1}{phv}{m}{sl} \color{Black} 
\\IMT Lucca, Bell Labs, ISTI-CNR Pisa 

\par\end{flushleft}\HorRule} 

\date{} 

\begin{document}

\maketitle 

\thispagestyle{fancy} 

\textit{\textbf{Main reference}}:  Coletto, M., Aiello, L. M., Lucchese, C., Silvestri, F. (2016) \textit{``On the Behaviour of Deviant Communities in Online Social Networks.''} Proceedings of the 10th International AAAI Conference on Web and Social Media. ICWSM 2016, May 17-20, Cologne, Germany.


\initial{T}\textbf{he structure of a social network is fundamentally related to the interests of its members. People assort spontaneously based on the topics that are relevant to them, forming social groups that revolve around different subjects. Online social media are also favorable ecosystems for the formation of topical communities centered on matters that are not commonly taken up by the general public because of the embarrassment, discomfort, or shock they may cause.  Those are communities that depict or discuss what are usually referred to as \textit{deviant behaviors}~\cite{clinard15sociology}, conducts that are commonly considered inappropriate because they are somehow violative of society's norms or moral standards that are shared among the majority of the members of society. Pornography consumption, drug use, excessive drinking, illegal hunting, eating disorders, or any self-harming or addictive practice are all examples of deviant behaviors.}

Many of them are represented, to different extents, on social media~\cite{haas10online,morgan10image,dechoundhury15anorexia}. However, since all these topics touch upon different societal taboos, the common-sense assumption is that they are embodied either in niche, isolated social groups or in communities that might be quite numerous but whose activity runs separately from the mainstream social media life. In line with this belief, also research has mostly considered those groups in isolation, focusing predominantly on the patterns of communications among community members~\cite{gareth15people} or, from a sociological perspective, on the motivations of their members and on the impact of the group activities on their lives and perceptions~\cite{attwood05people}.

In reality, people who are involved in deviant practices are not segregated outcasts, but are part of the fabric of the global society. As such, they can be members of multiple communities and interact with very diverse sets of people, possibly exposing their deviant behavior to the public. Here we focus on the deviant behaviour of \textit{adult content} consumption and analyze \textit{in context} the behavior of groups that post pornographic material online. Public depiction of pornographic material is considered inappropriate in most cultures, yet the number of consumers is strikingly high~\cite{sabina08nature}. Despite that, we are not aware of any study about online communities that produce that type of content interfaces with the rest of the social network. In this scenario, we contribute to shed light on three questions that are relevant to both network science and social sciences:
\begin{itemize}
\item How much pornographic content producers are secluded from the rest of the social network?
\item To what extent pornographic content spreads along the social ties?
\item What is the demographic composition of producers and consumers of adult content?
\end{itemize}

We analyze two large and \textit{fully anonymized} datasets sampled from Tumblr (130M users, 7B social links) and Flickr (39M users, 600M social links), two popular online platforms for microblogging and photo sharing, respectively. Consumption and production of adult content on general-purpose online social networks have never been studied at this scale before. We model three types of social interactions: \textit{i)} \textit{following} users enables a to receive updates from their content stream; \textit{ii)} \textit{liking} any piece of content produced by others expresses explicit interest in the activity of others; \textit{iii)} \textit{sharing} content produced by others increases the visibility of an item by re-posting it on the sharer's feed (Tumblr only).
 



We call deviant network the subgroup of all active users of the considered online social networks who published original pornographic content in their profiles  and their activities. We call these profiles \textit{\textbf{producers}}. A user profile is classified as producer if it is hit by a large number of search engine queries containing pornographic keywords, which is a strong indicator of the presence of adult content on the page. We call \textit{\textbf{consumers}} users who do not post new pornographic content, but they follow the producers and they eventually share and like their content.  This experimental setup led us to three main findings.

\vspace{6pt} \noindent \textbf{The community of adult content producers is small and clustered in dense sub-communities.} \vspace{3pt}
	
\begin{figure}[tp]
\includegraphics[clip=true, width=\columnwidth]{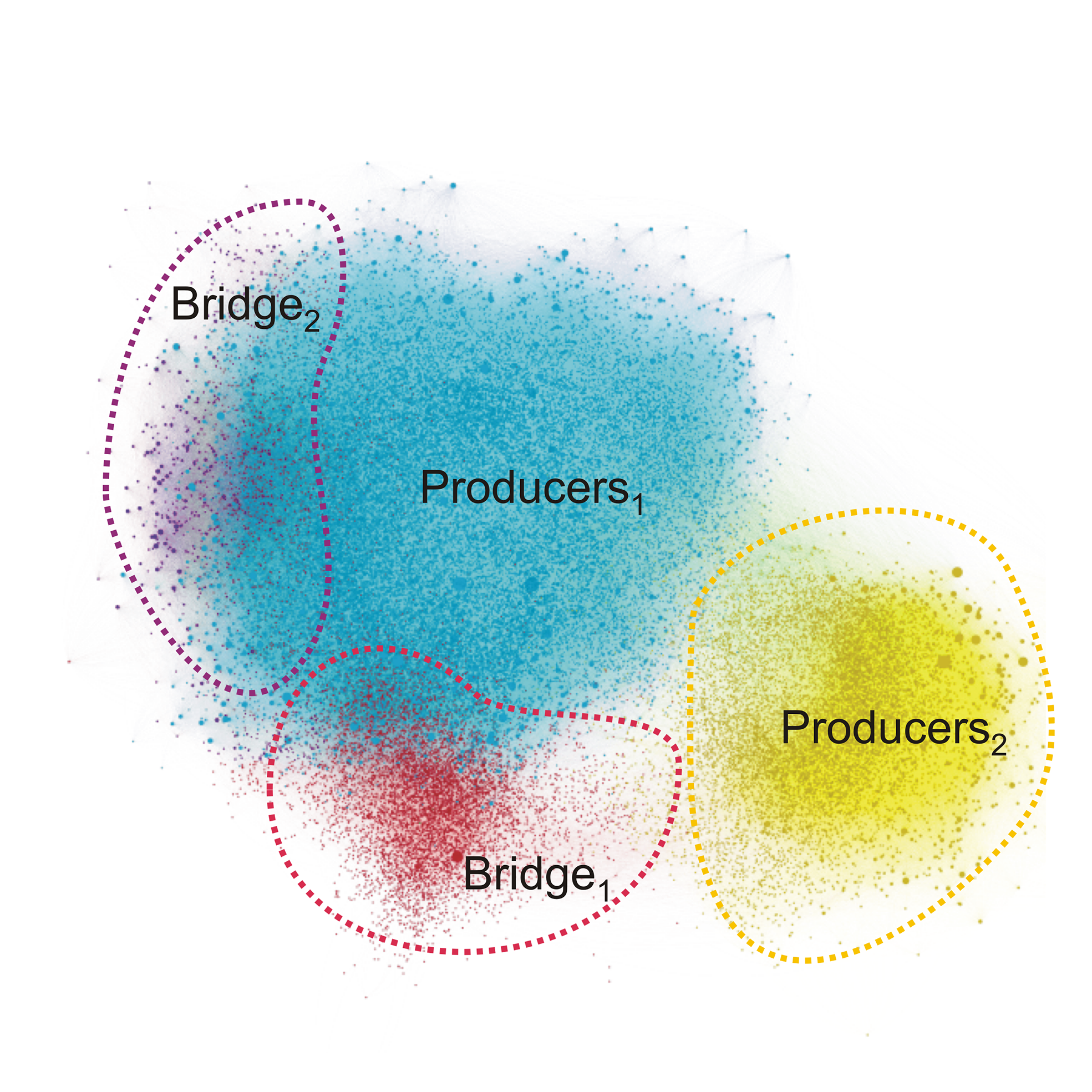}
\includegraphics[clip=true, width=\columnwidth]{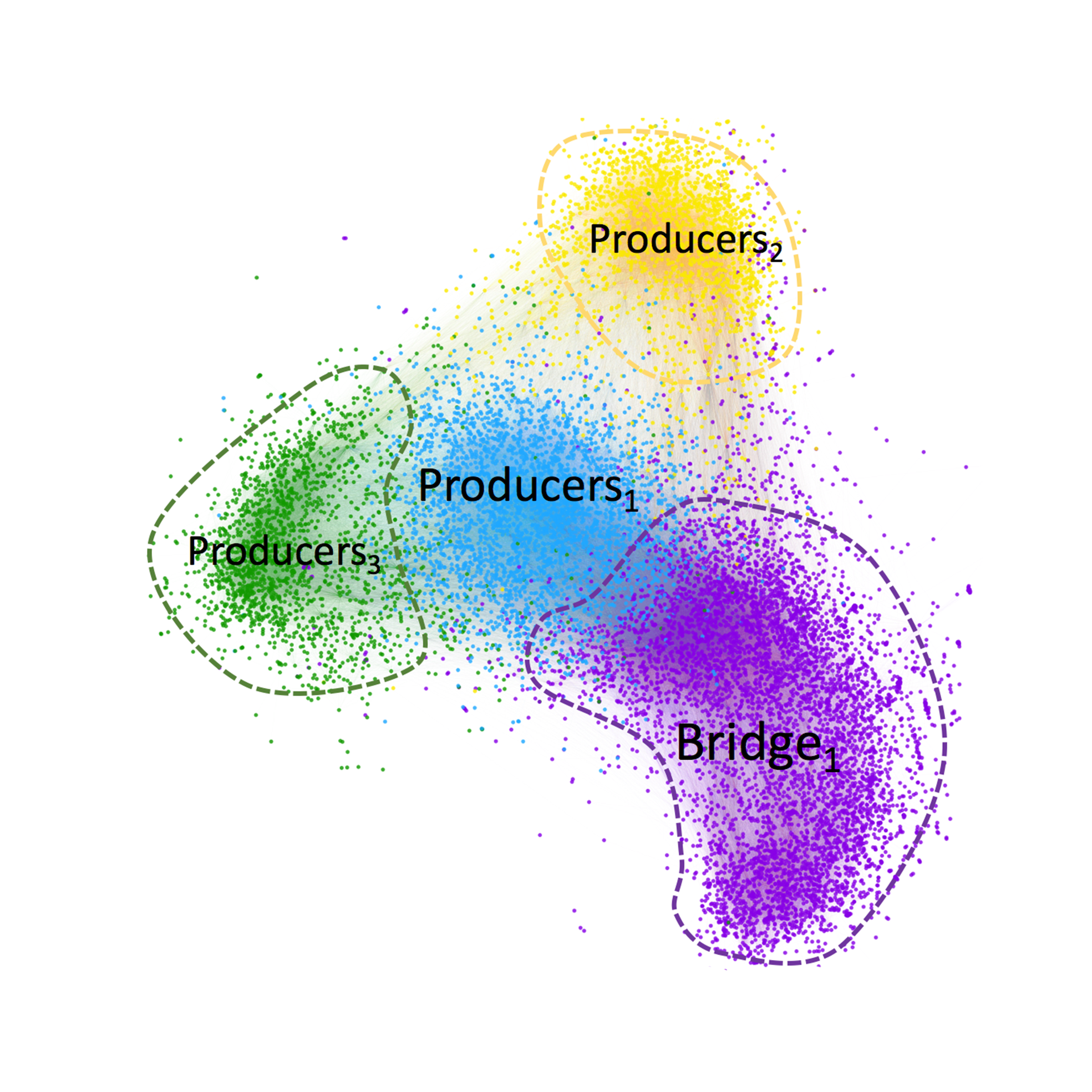}
\caption{Bird-eye view of the deviant network for Tumblr (reblog network, top) and Flickr (favorite network, bottom) with colors and labels denoting algorithmically-extracted communities. Each point is a user and the links represent their sharing activities.}
\label{fig:network_map}
\end{figure}

\noindent The deviant network is a tiny portion of the whole graph, representing about 0.8\% of all the users in the reblog graph in Tumblr, 1.1\% of all the nodes in the favorite graph in Flickr and a even smaller portion in the follow network (0.1\% in Tumblr and 0.4\% in Flickr). This tightly connected community is structured in subgroups, but it is linked with the rest of the network with a very high number of (mostly incoming) ties. In Figure~\ref{fig:network_map} we show the internal structure of the deviant network. In Tumblr more than $90\%$ of all the blogs in the two largest clusters contain blogs that \textit{exclusively} produce explicit adult content, aimed at an heterosexual public (\textit{Producers}$_1$) or at a male homosexual public (\textit{Producers}$_2$). The blogs in the two remaining communities post less explicit adult content and more sporadically, often by means of reblogging. They either focus on celebrities (\textit{Bridge}$_1$), or function as aggregator blogs with high content variety, including depiction of nudity (\textit{Bridge}$_2$). In Flickr the two main \textit{Producers} clusters are similar to Tumblr's, in terms of type of content posted. In addition to those, we find a third cluster (\textit{Producers}$_3$) producing predominantly content related to the transvestite and transsexual communities. Producers in other sub-communities post less explicit content (soft porn, artistic nudity, manga); we call these groups \textbf{\textit{bridge}} communities as their main focus is often not on deviant content and they act also as link towards the rest of the graph.
	
\vspace{6pt} \noindent \textbf{The pornographic content spreads widely in the network.} \vspace{3pt}

\begin{figure}[tp]
\centering
\includegraphics[clip=true, width=\columnwidth]{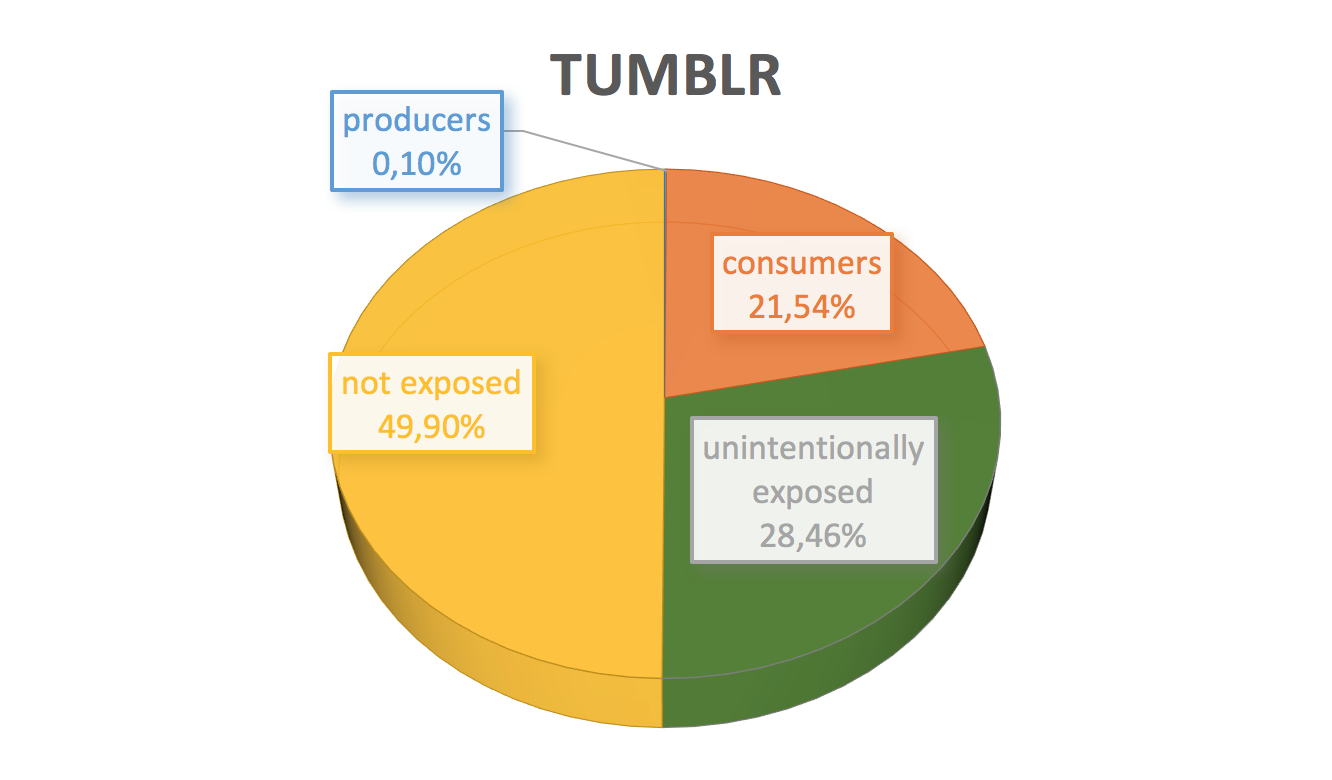}
\includegraphics[clip=true, width=\columnwidth]{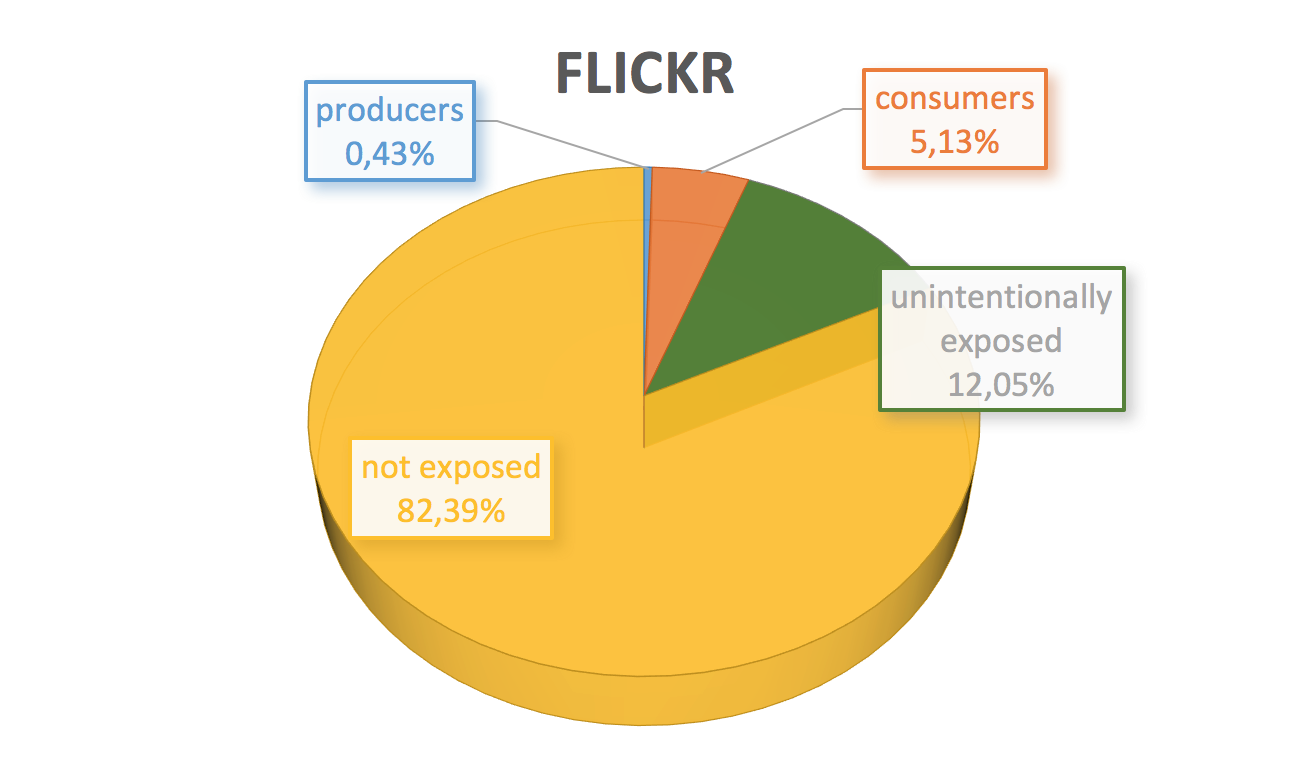}
\caption{User distribution for Tumblr (top) and Flickr (bottom). }
\label{fig:torta}
\end{figure}

Even though the vast majority of pornographic material is posted by a very small core of users, the adult content reaches a much wider audience, spreading through social ties. In Figure~\ref{fig:torta} we show the distribution of the users among adult content producers, \textit{\textbf{consumers}} (i.e., people who like or reblog pornographic material originally posted by others or follow a producer) and  \textit{\textbf{unintentionally exposed}} users who do not follow any producer and do not reblog their content, but happen to follow at least one profile who pushes adult content in their feed through reblogging. In Flickr, we consider unintentionally exposed all the users who follow people who liked adult content; this choice is motivated by the fact that the Flickr feed might show pictures that your social contacts recently liked.
In Tumblr, adult content consumers are 22\% of our sample; 5\% in Flickr. In Tumblr the spreading is wider also because the platform enables sharing actions (reblog) which are not present in Flickr. Moreover, although the consumption of deviant content remains a minority behavior, the average local perception of users is that neighboring nodes reblog more deviant content than they do.
	
\vspace{6pt} \noindent \textbf{No, pornography is not for boys only. Peak of consumption is at 50-55 years for males and at 20-25 for females}\vspace{3pt}

\begin{figure}[tp]
\centering
\includegraphics[clip=true, width=\columnwidth]{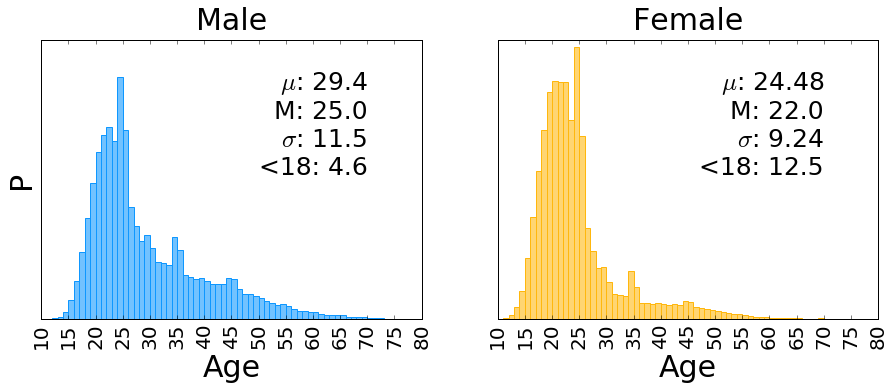}
\includegraphics[clip=true, width=\columnwidth]{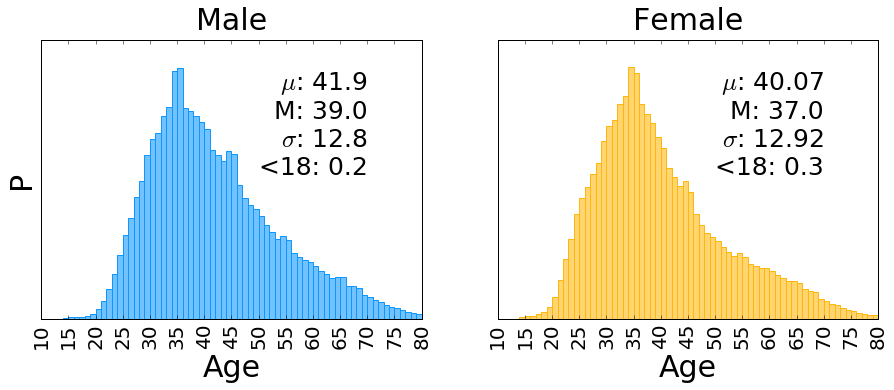}
\caption{Age distribution of users in our dataset for Tumblr (top) and Flickr (bottom). Mean $\mu$, median M, standard deviation $\sigma$, and percentage of users under 18 years old are reported.}
\label{fig:age_total}
\end{figure}

Tumblr is used predominantly by young females: based on a statistics over 1.7M Tumblr users who self-reported their demographic information, we estimate that the average user age is 26 and $72\%$ of the users are female. On a sample of 12.3M Flickr users with demographic information, we find that the platform is used mostly by adult males: the average age is 41 and around $59\%$ of users are male (Figure~\ref{fig:age_total}).

\begin{figure*}[tp]
\centering
\includegraphics[clip=true, width=.90\textwidth]{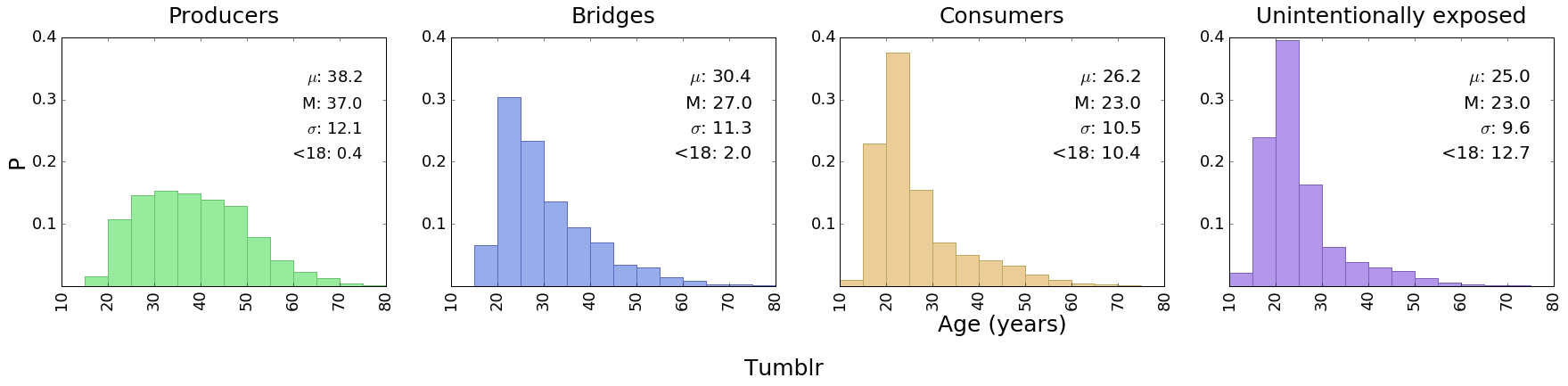}\\
\includegraphics[clip=true, width=.90\textwidth]{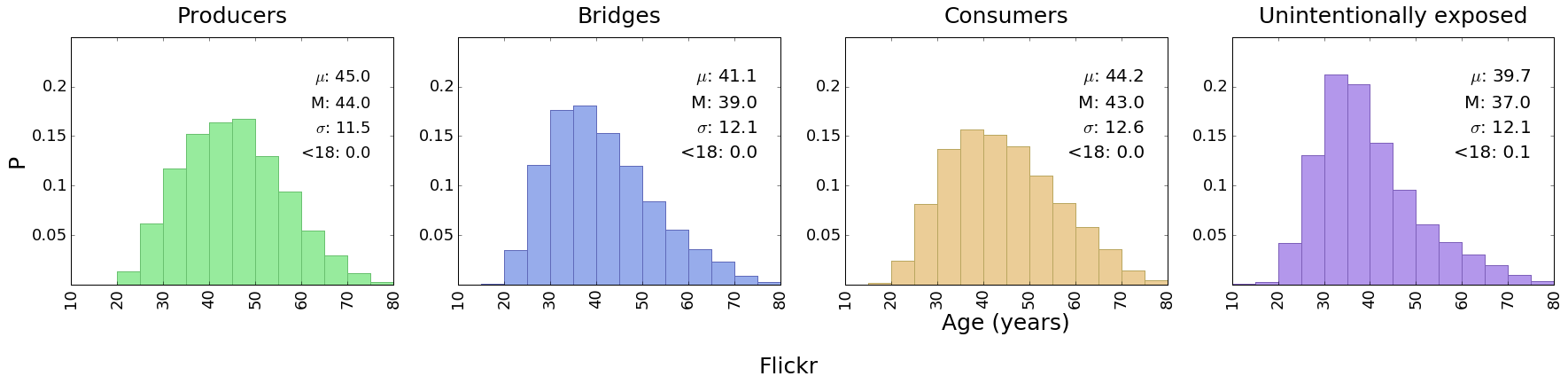}
\caption{Age distribution of different groups of producers and consumers of adult content in Tumblr (top) and in Flickr (bottom). Average ($\mu$), median (M), standard deviation ($\sigma$), and percentage of underage users ($<18$) are reported.}
\label{fig:age_group}
\end{figure*}

In Figure~\ref{fig:age_group} we report the age distribution for users at different levels of exposure to adult content. Active consumers are users who share (in Tumblr) or like (in Flickr) adult content, passive consumers are users who follow adult content producers. There are clear differences in the age distributions between producers and consumers of adult content that are compatible with previous literature on use of adult material, with producers being sensibly older than other gorups. We find that a non-negligible large fraction of underage users can be inadvertently exposed to such content through indirect links.
	
\begin{figure}[tp]
\centering
\includegraphics[clip=true, width=\columnwidth]{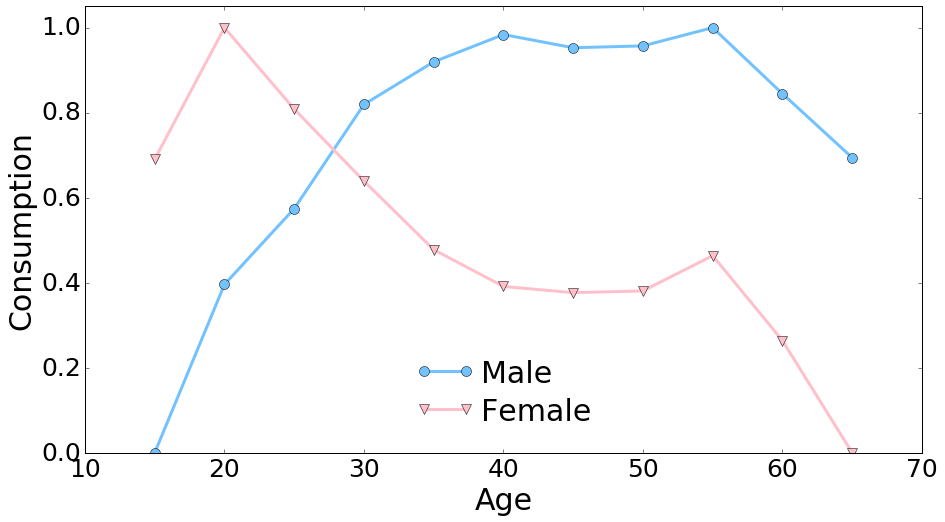}
\includegraphics[clip=true, width=\columnwidth]{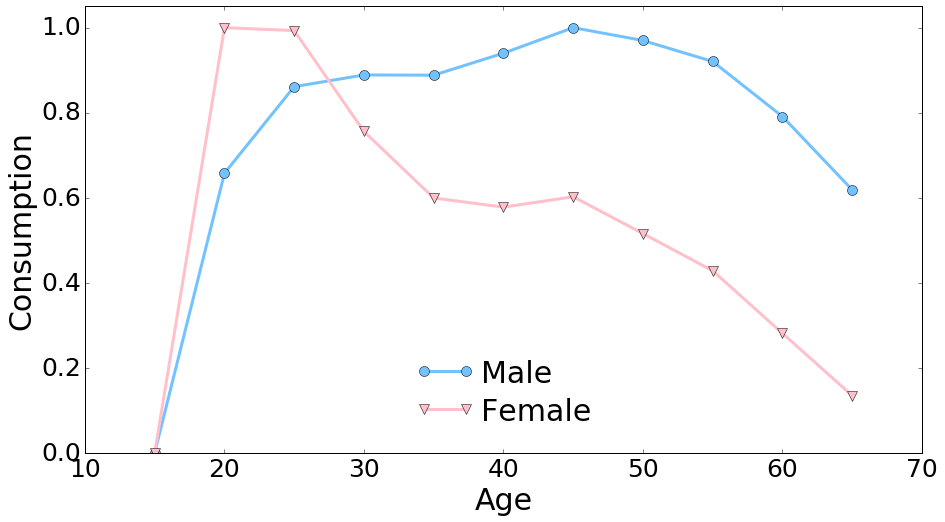}
\caption{Ratio of male and female consuming adult content for different age bands in Tumblr (top) and in Flickr (bottom). Each line in the plot is min-max normalized.}
\label{fig:age_ratio}
\end{figure}

It is common sense that males consume more pornography than female. We find that people in the producer class are mostly male ($84\%$ in Tumblr and $76\%$ in Flickr), but on aggregate the gender distribution of consumers reflects the gender distribution in the overall population, supporting the intuition that gender does not play a major role in the practice of pornography consumption as one would expect. However, differences emerge when age is factored in. The consumption of adult content is substantially balanced between the two genders under 25 years old. After that age, the percentage of male users increases progressively compared to the female users. In Figure~\ref{fig:age_ratio} we report the relative consumption trend for different age ranges for both genders; 0/1 indicate the minimum/maximum level of consumption for a specific gender among all age bands). The consumption rises steadily for males and peaks between age 40 to 55.  In contrast, women, although less exposed than men at any age, have their peak in their 20s, much earlier than men, and then drop rapidly. Interestingly enough, the trends are consistent across the two platforms.



\bibliographystyle{alpha}
\small
\bibliography{bibliography}



\end{document}